\documentclass[12pt]{article}
\usepackage{a4wide}
\usepackage{amssymb}
\usepackage{graphicx}
\begin{document}
{\renewcommand{\thefootnote}{\fnsymbol{footnote}}
\hfill  IGPG--08/2--6\\
\medskip
\begin{center}
{\LARGE Loop Quantum Cosmology:\\ Effective theories and oscillating
universes\footnote{Chapter contributed to: ``Beyond the Big Bang'',
Ed.: R\"udiger Vaas (Springer Verlag, 2008)}}\\
\vspace{1.5em}
Martin Bojowald
and Reza Tavakol
\\
\vspace{0.5em}
$^1$Institute for Gravitational Physics and Geometry,
The Pennsylvania State
University,\\
104 Davey Lab, University Park, PA 16802, USA\\
\vspace{0.5em}
$^2$ Astronomy Unit, School of Mathematical Sciences,
Queen Mary, University of London, 
Mile End Road, London, E1 4NS, UK

\vspace{1.5em}

\end{center}
}

\setcounter{footnote}{0}
\newcommand{\lpl}{\ell_{\rm P}}

\begin{abstract}
Despite its great successes in accounting for the current
observations, the so called `standard' model of cosmology faces a
number of fundamental unresolved questions. Paramount among these are
those relating to the nature of the origin of the universe and its
early evolution. Regarding the question of origin, the main difficulty
has been the fact that within the classical general relativistic
framework, the `origin' is almost always a singular event at which the
laws of physics break down, thus making it impossible for such an
event, or epochs prior to it, to be studied.  Recent studies have
shown that Loop Quantum Cosmology may provide a non-singular framework
where these questions can be addressed. The crucial role here is
played by quantum effects, i.e.\ corrections to the classical
equations of motion, which are incorporated in effective equations
employed to develop cosmological scenarios.

In this chapter we shall consider the three main types of quantum
effects expected to be present within such a framework and discuss
some of their consequences for the effective equations. In particular
we discuss how such corrections can allow the construction of
non-singular emergent scenarios for the origin of the universe, which
are past-eternal, oscillating and naturally emerge into an
inflationary phase.  These scenarios provide a physically plausible
picture for the origin and early phases of the universe, which is in
principle testable.  We pay special attention to the interplay between
these different types of correction terms. Given the absence, so far,
of a complete derivation of such corrections in general settings, it
is important to bear in mind the questions of consistency and
robustness of scenarios based on partial inclusion of such effects.
\end{abstract}

\newpage

\section{Introduction}
Cosmology has undergone tremendous advances in recent years, both
theoretically and observationally.  An important outcome of these
developments has been the emergence of a `standard' model of
cosmology, with an early phase of accelerated expansion (the so called
inflationary phase) followed by a Hot Big Bang expansion phase, which
very successfully describes the central features observed by recent
high-precision observations. Despite these successes, however, a
number of fundamental questions remain. Foremost among these is the
unsatisfactory singular origin for the universe predicted by this
model, where the laws of physics break down and scientific
predictability comes to an end. Another crucial shortcoming is the
absence of a successful model of inflation that is properly situated
within a fundamental theory of quantum gravity. It has long been hoped
that these problems can be successfully resolved once a complete
theory of quantum gravity is known. In this chapter we shall
concentrate on some recent developments in Loop Quantum Cosmology
(LQC), i.e.  the specialisation of Loop Quantum Gravity to
cosmological backgrounds, which potentially have important
consequences for these questions.

Within the framework of LQC (see \cite{LivRev} for a detailed review)
space-time is fundamentally described by a discrete structure whose
dynamics is governed by difference rather than differential
equations. This discreteness is an essential feature on small scales,
such as those prevailing in the earliest phases of the universe.
According to LQC, as the universe evolves and grows in size, this
initial discrete quantum phase is succeeded by a tamer semi-classical
phase in which the spacetime is well approximated by classical
geometry, but governed by dynamical equations which are corrected by
quantisation effects \cite{Inflation}.  Finally as the universe grows
still larger, the familiar classical continuum picture is approached
to a good approximation.

To get a glimpse of possible LQC effects that are likely to be
present, we consider the standard isotropic and homogeneous
Friedmann-Lema\^{\i}tre-Robertson-Walker (FLRW) cosmological models
which are sourced by a scalar field $\phi$ with a potential $V (\phi)$
and a Hamiltonian ${H}_{\phi}=\frac{1}{2}a^{-3}p_{\phi}^2+a^3V(\phi)$,
where $a$ is the scale factor and $p_{\phi}$ is the scalar field
momentum.  The scalar field is a common choice for the `matter' source
in studies of the early universe, since once inflation begins all
other matter sources except the scalar field are rapidly redshifted
away.  Classically the dynamics is described by the Hamiltonian
constraint $H=0 $ which can be written as
\begin{equation}
\label{hamil} 
 \frac{1}{2}a^{-3}p_{\phi}^2+a^3V(\phi)=\frac{3c^2}{8\pi
 G}(\dot{a}^2+c^2k)a\,.
\end{equation}
Here, $k$ is the curvature parameter, which is zero for a spatially
flat universe and takes values $+1$ and $-1$ for positively and
negatively curved universes.  In the homogeneous and isotropic case
considered here this constraint equation reduces to the familiar
Friedmann equation (which is the relevant Einstein equation in this
case).  Thus this constraint relates the matter (here taken to be the
scalar field) Hamiltonian, given by the terms on the left hand side of
equation (\ref{hamil}), to the time derivative of the scale factor on
the right hand side of this equation.

An inspection of this equation readily 
demonstrates the difficulty usually faced in the 
classical domain, namely the fact that solutions to this equation 
are singular, all reaching a zero volume as $a\to 0$,
at some point in the past (in all isotropic models)
or both future and past (in the case of closed models
with $k=+1$). At such points
the matter Hamiltonian (energy) diverges ($H_{\phi}\to\infty$),
while at the same time $\dot{a}\to\pm\infty$. 
This represents a space-time singularity which occurs for 
typical cosmological solutions in general relativity 
\cite{HawkingEllis,Senovilla}. The presence of such singularities
has posed a severe barrier to the formulation of physically meaningful
cosmological models within the context of classical general relativity,
since at such points the theory itself 
breaks down and fails to be applicable. Thus general relativity cannot
tell us what happens at, or beyond, such a singular point. 

It has long been hoped that this barrier would be removed once a
complete description is available which involves the quantisation of
general relativity. The basic idea is that once the universe shrinks
to a small enough size, close to the Planck length $\ell_{\rm
P}=\sqrt{G\hbar/c^3}$ (a fundamental length scale constructed in terms
of the constants of nature $G$, Planck's constant $\hbar$ and the
speed of light $c$), the quantisation would result in quantum
corrections to the classical cosmological evolution equations which
could dramatically change their behaviours and in particular prevent
such singularities from forming.

An important outcome of the developments in LQC has been the
demonstration that there are indeed several different quantum gravity
effects, with their corresponding quantum corrections, which could in
principle result in well behaved singularity free cosmological
scenarios, with potentially observable consequences.  In the following
we shall briefly review these effects and discuss some of their
consequences for non-singular cosmological model construction. This
will be presented from a general viewpoint: We do not focus on precise
results in very specific models, but rather discuss qualitative
aspects expected more generally.
\section{Quantum corrections}
The Friedmann equation (\ref{hamil}) already highlights two sources of
singular behaviour which need to be remedied by quantum corrections
from loop quantum cosmology, if we are to obtain a non-singular model.
These are the kinetic term, $\frac{1}{2}a^{-3}p_{\phi}^2$, which
clearly diverges as $a\to0$ and the term $\dot{a}$, on the right hand
side of this equation, which also diverges at a cosmological
singularity.

It is important to recall that these two particular sources of
classical singularities are directly related to the potential blow up
of the curvature (a function of $\dot{a}$ and $1/a$ in an isotropic
and homogeneous geometry) which is the hallmark of singularities in
the well known solutions of general relativity. This observation may
suggest that only quantum corrections relating to terms which
correspond to classical unbounded curvature might be the ones which
are of relevance in preventing classical singularities to form. But
the celebrated singularity theorems do not even refer to curvature
blow-ups, indicating that such a restriction is likely to be
misleading. In a general unbiased approach, all possible quantum
correction terms must be considered, which are to be derived from the
underlying quantum equations and evaluated in cosmological or black
hole scenarios. Only then can one be certain of addressing the problem
of classical singularities which arise within the classical
relativistic framework in a general and systematic way. In this
chapter we present a summary of the present status of these questions.

To proceed we note that LQC corrections do indeed modify the two terms
mentioned above, which in this case are responsible for the curvature
blow-up. First, the inverse volume $a^{-3}$, when adequately
quantised, acquires quantum corrections which become significant as
$a$ becomes small. In fact it turns out that in the case of exactly
homogeneous and isotropic models considered here, the classical
divergence is cut off by quantum effects and this expression remains
finite. (We note that in the isotropic case there are cancellations
which render the right hand side of Eq.~(\ref{hamil}) free of inverse
powers \cite{Closed,AmbigConstr}. This will, however, no longer be the
case in less symmetric cases which have corrections of the same type
as in the isotropic matter part.)  Secondly, in loop quantum
cosmology, the term $\dot{a}^2$ on the right hand side of
Eq.~(\ref{hamil}) appears only as the leading term in a power series
expansion of a more complicated function of $\dot{a}$. On classical
scales, where general relativity is currently probed, $\dot{a}$ is
small and the leading term is the only relevant one. However, as the
universe approaches its classical singularity, $\dot{a}$ increases
without bounds making the higher order terms noticeable at some
point. It turns out that this results in a power series in $\dot{a}$,
which has $\dot{a}^2$ as its leading term, and which sums to a bounded
function which can cut off the potential divergences.

These two effects have already been explored in a number of
phenomenological models, some of which we shall describe below.  In
addition to these effects, however, there is a third, potentially more
important, source of corrections which is not so easy to visualise or
to derive systematically. According to general considerations of
effective actions, such corrections in the isotropic and homogeneous
settings are expected to occur as higher derivatives of the scale
factor (such as $\ddot{a}$, \ldots) in the Friedmann equation.  The
inclusion of these corrections is crucial for a reliable derivation of
effective equations which need to include the complete set of quantum
corrections. Thus a full consideration of loop quantum corrections
would, in addition to inverse volume corrections, also include higher
powers as well as higher derivatives of $\dot{a}$.  Since these extra
terms are also divergent at a cosmological singularity, they must be
included in any comprehensive analysis.  In the following we shall
briefly discuss these three sources of quantum corrections to the
effective equations.
\subsection{Inverse volume effects}
In the context of homogeneous and isotropic cosmological models, the
main reason for the occurrence of classical singularities is the
presence of the divergent inverse volume term in the matter
Hamiltonian. An outcome of isotropic LQC, based on general techniques
developed in loop quantum gravity \cite{QSDI,QSDV}, has been to show
that any well-defined quantization of the classical inverse volume is
finite \cite{InvScale}. (Although the case of anisotropic models is
different, the differences are not problematic and are well-understood
\cite{DegFull}.)  In LQC settings, the effective equations do not
include the inverse volume operator itself but only expectation values
or the corresponding eigenvalue function, which can be well
approximated by a continuous function $D(q)a^{-3}$, to replace
$a^{-3}$, where the function \cite{Ambig,ICGC}
\begin{eqnarray} \label{D}
D(q) &=&
\left\{\frac{3}{2l}q^{1-l}\left[(l+2)^{-1}
\left((q+1)^{l+2}-|q-1|^{l+2}\right)\right.\right. \nonumber\\
 &-& \left.\left.\frac{1}{1 + l}q
\left((q+1)^{l+1}-{\rm
sgn}(q-1)
|q-1|^{l+1}\right)\right]\right\}^{3/(2-2l)}
\end{eqnarray}
is expressed in terms of the dimensionless ratio $q=a^2/a_*^2$. Here,
$a_*^2=\ell_{\rm P}^2j/3$ and $j$ and $l$ are parameters that arise in
the process of quantisation.  An important feature of this
quantisation scheme is that these parameters are not uniquely
specified: $l$ lies in the range zero to one and $j$ takes
half-integer values.  It turns out, however, that both these
parameters would be more restricted if analogous operators were
derived without assuming symmetries. Current constraints on $l$ are
$l_n=1-1/2n$ where $n$ takes integer values \cite{Robust}.  The
restrictions on $j$ are less clear since its value depends not only on
how the operator itself is derived but also on the precise form of
states used to implement symmetries
\cite{InhomLattice,SchwarzN}. (Arguments for $j=1/2$ have been put
forward \cite{AmbigConstr,AlexAmbig}, but nevertheless larger values
of $j$ can occur in homogeneous models to capture more general
inhomogeneous features \cite{SchwarzN}.)  It is therefore important to
bear in mind that the quantisation procedures employed involve
ambiguities due in part to non-uniqueness of such parameters.
\begin{figure}
\resizebox{0.9\textwidth}{!}{%
  \includegraphics{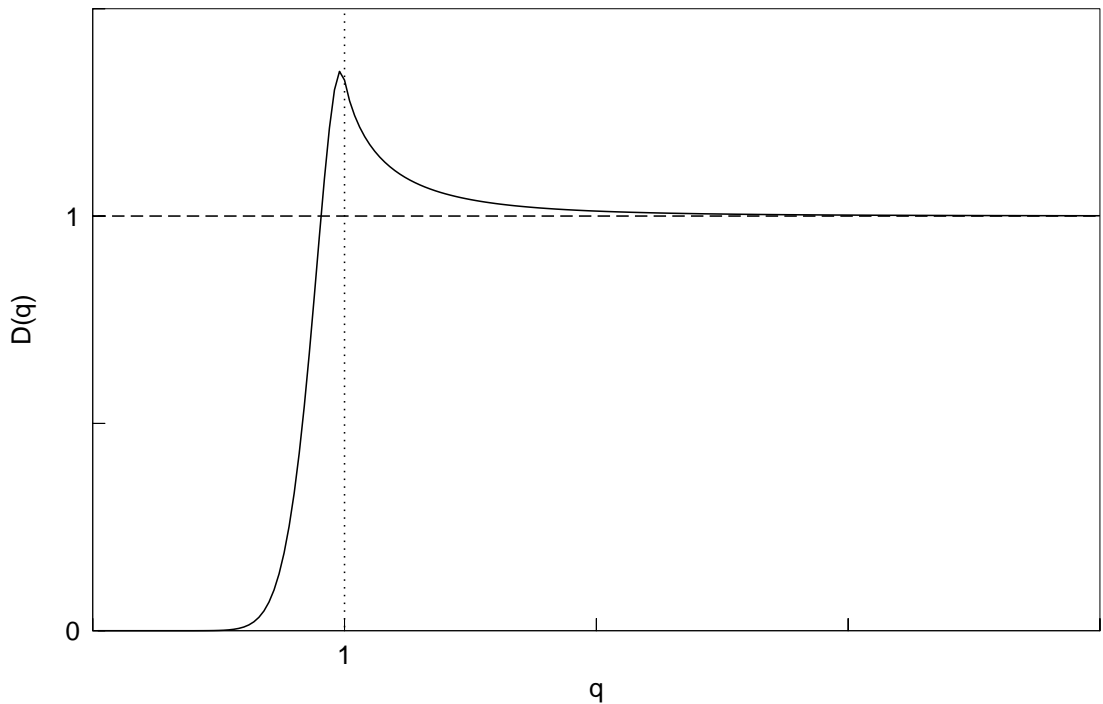}}
\caption{\label{Fig:D} {In isotropic loop quantum cosmology, functions
with inverse powers of the scale factor are finite instead of
diverging at $a=0$ as happens classically. This is captured in a
correction function $D(q)$ where $q=a^2/a_*^2$ is defined in terms of
a characteristic scale $a_*$ due to quantum gravity. A plot of the
function $D$ against $q$ shows how $D$ and hence the quantised inverse
volume deviate from the classical result (dashed) and approach zero as
$a \to 0$. }}
\end{figure}
The function $D$ codifies the semi-classical corrections to the
evolution equations and has a number of important qualitative
features, as can be seen from Fig.~\ref{Fig:D}.  It tends to zero as
$a \to 0$, increases for small values of $a$, peaks at intermediate
values of $a$ ($a\sim a_*$) and as $a$ increases it eventually
approaches the classical expectation value of one from above.  
(Note that a fixed $a_*$ in combination with the rescaling freedom of
$a$ in flat models induced by changing spatial coordinates is only an
apparent problem which is solved when viewed in the appropriate
inhomogeneous context; see \cite{InhomLattice}. The rescaling freedom
is simply an accidental property of specific classical solutions which
is broken by quantum effects (as it would be by spatial curvature or
inhomogeneities). The new parameter $a_*$ is then related to the
underlying scale of the discrete quantum gravity state responsible for
the breaking of rescaling freedom.)

Thus one effect of quantum corrections on the evolution
Eq.~(\ref{hamil}) is to replace the inverse volume in this equation by
$D(q)a^{-3}$.  We shall see that these features induce radically
different behaviours in LQC models compared to the corresponding
classical ones.  A simple way to see some of the qualitative effects
of these modifications is to rewrite the quantum corrected equations
in the form of standard Einstein equations in terms of an effective
perfect fluid with a variable equation of state \cite{Oscill} (see
also \cite{EffHam}). In that case the increasing behaviour of $D(q)$
for small values of $a$ has the consequence of making the effective
matter Hamiltonian an increasing function of the scale factor.  Thus
for small values of $a$ the energy increases with volume, implying
that in these regimes the pressure of the fluid, defined as the
negative derivative of energy with respect to volume, is negative. The
presence of such a negative pressure could in principle provide a
mechanism for resisting the collapse of the universe into a
singularity. As we shall see below, this picture has indeed been shown
to occur in a number of isotropic scenarios where the inverse volume
correction has been included.  This correction on its own, however, is
not sufficient to ensure singularity avoidance in more general
settings.  In such cases other corrections, such as those discussed
below, are also likely to be required.

\subsection{Higher order corrections}
In homogeneous and isotropic models inverse volume effects occur
mainly (but not only) in the matter Hamiltonians.  As was mentioned
above, it is expected on general grounds that in addition to these
effects there also exist higher order curvature corrections.  (See
\cite{SemiClass}; such terms have also been computed by a WKB analysis
\cite{EffHam,DiscCorr}.) In the case of isotropic models these terms
take the form of a subseries of higher powers of $\dot{a}$, i.e.\
$\sum_nc_n \dot{a}^n$, with coefficients $c_n$ which may depend on $a$
(but not on derivatives). Although different possibilities exist
depending on the precise quantization, the qualitative form is known
to be $a^{-2}\dot{a}^2(1+O(\ell_{\rm P}^2\dot{a}^2/a^2c^2))$ which,
for a specific example, sums to a closed form function
\begin{equation} \label{higher}
\frac{c^2}{2\ell_{\rm P}^2}\left(1-\sqrt{1-\frac{4\ell_{\rm
P}^2\dot{a}^2}{a^2c^2}}\right) \sim \frac{\dot{a}^2}{a^2}
-\frac{\ell_{\rm P}^2}{c^2}\frac{\dot{a}^4}{a^4}+\cdots
\end{equation}
(see also Sec.~\ref{s:QuantState}).  Despite the remaining
quantization ambiguities, qualitative properties are quite robust and
follow from very general considerations in the full setting of loop
quantum gravity which allow only specific types of variables, namely
holonomies and fluxes, to be quantized in a way which allows the
general covariance and background independence of general relativity
to be realized after quantization.
In our case, the form of this function replacing the
classical $\dot{a}^2/a^2$ already provides a way of seeing how such
quantum corrections can lead to the boundedness of the curvature, by
noting that the reality of the above expression imposes an upper limit
on the curvature related term $\dot{a}^2$.  The coefficients $c_n$ of
the terms $c_n\dot{a}^n$ in this subseries, most of which are
exceedingly small yet important for the final result, can in principle
be determined exactly from the quantum theory.  Note, however, that
the precise bounded function requires the taking into account of all
the infinite number of higher order terms in the subseries.

An important point to bear in mind in this connection
is that there are additional quantum corrections, discussed in the
following subsection, which are likely to be larger than almost all
these terms in the series, independently of the regime being studied.
As a result the precise form of the above function, and in particular
its ability to provide an upper bound for curvature, is reliable only
if one can show that all these other quantum corrections are
absent. Otherwise, the analysis needs to be repeated by including all
these other terms.

\subsection{Higher derivative corrections}
\label{s:Higher}
A potentially far more important set of 
corrections concerns terms which involve 
higher derivatives. In contrast to the two types of corrections
discussed above, terms of this type arise in the effective equations 
more indirectly. Intuitively, they can be understood as
capturing the essential dynamical behaviour of a quantum system
described by a wave function \cite{Karpacz}. Unlike a
classical mechanical system with a finite number of degrees of
freedom, a quantum mechanical state or wave function requires the
specification of infinitely many variables. In addition to expectation
values of basic operators $\hat{q}$ and $\hat{p}$, corresponding to
position and momentum, which in a
semiclassical state can be identified with the classical variables,
there are independent variables which represent the infinitely many
different ways the wave function can be deformed from a 
Gaussian form. These include 
the spread and various forms of distortions of the
wave function.

Fortunately, from a practical point of view, not all these infinite
number of variables are expected to be important observationally.
Thus in practice one can usually restrict considerations to a finite,
and small, number of such deformations.  It should, however, be borne
in mind that even when such variables are not observed directly, they
may still have important indirect dynamical effects, even on the
expectation values. As a wave packet moves, it generically spreads and
deforms in ways which also affect the evolution of the expectation
values. All these infinite number of quantum variables are in
principle dynamical and, via their back-reaction on the motion of what
can be considered as the classical variables, lead to quantum
corrections.

A simple way of illustrating this is by considering an an-harmonic
oscillator with the Hamiltonian operator
$\hat{H}=\frac{1}{2m}\hat{p}^2+V(\hat{q})= \frac{1}{2m}\hat{p}^2+
\frac{1}{2}m\omega^2\hat{q}^2+ \frac{1}{3}\lambda\hat{q}^3$, where $V$
is the potential, $m$ is the mass, $\omega$ is the frequency and
$\lambda$ is the nonlinear coupling constant.  The equation of motion
for the expectation value $\langle\hat{p}\rangle$ can then be written,
using the commutator
$[\hat{p},\hat{H}]=\hat{p}\hat{H}-\hat{H}\hat{p}$, as
\begin{eqnarray*}
 \frac{{\rm d}}{{\rm d} t}\langle\hat{p}\rangle &=&
\frac{1}{i\hbar}\langle[\hat{p},\hat{H}]\rangle=
-m\omega^2\langle\hat{q}\rangle-\lambda\langle\hat{q}^2\rangle=
-m\omega^2\langle\hat{q}\rangle-
\lambda\langle\hat{q}\rangle^2-\lambda\Delta q^2\\
&=&-V'(\langle\hat{q}\rangle)-\lambda\Delta q^2\,,
\end{eqnarray*}
where $\Delta
q=\sqrt{\langle\hat{q}^2\rangle-\langle\hat{q}\rangle^2}$.  As can be
seen, the classical equation of motion $\dot{p}=-V'(q)$ is changed by
the addition of an extra term, containing $\Delta q$, which would
classically be zero and corresponds to quantum fluctuations.
Moreover, the rate of change of $\Delta q$ with respect to time, ${\rm
d}\Delta q/{\rm d} t$, is itself non-zero as the quantum state spreads
during its evolution.  These variables, together with the infinite
number of others which characterize a wave function, are thus coupled
to each other. (A systematic approximation in adiabatic regimes where
fluctuations do not change rapidly reproduces low energy effective
actions \cite{EffAc,EffectiveEOM}.) This complicated back-reaction
effect is captured by correction terms in effective equations which,
from an effective action point of view, in general require higher time
derivatives.

The above discussion indicates that such higher derivative correction
terms are the most difficult to compute since, in a Hamiltonian
formulation, they require all the details of a dynamical semiclassical
quantum state through $\Delta q(t)$ and other quantum parameters. At
present, they are known precisely only for a free, massless scalar in
a spatially flat isotropic universe.  In itself this is a very simple
system which is not very interesting from a dynamical point of view.
Nevertheless, it is closely related to an exactly solvable system
\cite{BouncePert,BounceCohStates} which therefore allows it to be used
to set up a general perturbation theory. This can enable the study of
more realistic models by including a matter potential or
inhomogeneities and proceeding perturbatively to obtain the resulting
higher derivative terms which are required for a complete effective
analysis.  This question is currently under investigation, with some
initial results described in Sec.~\ref{s:QuantState} below. Clearly
the construction of more realistic cosmological scenarios within loop
quantum cosmology, which would require the inclusion of such quantum
back-reaction terms (to be distinguished from classical back-reaction
of developing inhomogeneities), will need to await the outcome of
these investigations.
\subsection{Combined effects of multiple corrections}
The above discussions highlight the fact that the derivation of full
effective equations within the loop quantum cosmology framework is
likely to include three different types of quantum corrections.  These
corrections fall into two categories. First, there are those that are
direct consequences of quantum geometry and already occur in static
settings {for which no equations of motion need to be solved.}  The
inverse volume and higher degree corrections discussed above belong to
this category. On the other hand the higher derivative terms, which
are due to genuine quantum effects and constitute the most important
corrections, occur in generic dynamical quantum systems. While effects
in the first category become important only when the universe enters
high curvature or strong quantum regimes, and subside quickly once it
exits such regimes, effects belonging to the second category are
active at all times. Even when they are tiny, they can still add up
during the long evolution times which are the hallmark of cosmological
scenarios, and this cumulative effect can change the dynamical
behaviour of the universe dramatically.

Clearly the construction of any realistic cosmological scenario needs
to include all these corrections simultaneously.  An important issue
to bear in mind in this connection is that the relative sizes of these
effects are regime dependent, and in general largely unknown at
present. As a result, so far the consequences of different loop
quantum cosmological corrections have often been studied individually
by judicious choices of suitable parameters which single out a
specific correction term as the dominant one. For instance, inverse
volume effects can be studied in isolation by choosing a large value
for the parameter $j$ in the function $D(q)$, which has the effect of
shifting the peak in Fig.~\ref{Fig:D} to larger values of $a$,
corresponding to larger volumes. In this way, inverse volume
corrections will still be relevant even for relatively large
universes, where the other curvature corrections can be
ignored. Similarly, the effects of higher order corrections can be
studied in isolation by including a free scalar field as the only
matter source and choosing its energy, determined by $p_{\phi}$, to be
very large
\cite{QuantumBigBang,APS}. An analogous procedure, however, is not
currently known for genuine quantum effects (higher derivative terms)
since they are quantum fluctuations which are determined by the
quantum dynamics and cannot be tuned easily. The first two effects,
however, can be and have been studied in this way, revealing possible
physical effects concerning the questions of origin and early
inflationary phases of the universe.

In general, therefore, given this multiplicity of effects, care must
be taken in constructing realistic cosmological scenarios within loop
quantum cosmology. In particular, two important questions need to be
borne in mind.  The first concerns the consistency of the set of
corrections taken into account. For example, often regimes which
require higher order correction terms would also necessitate the
inclusion of higher derivative terms.  Thus, unless higher derivative
terms can be shown to be absent, keeping only the higher order terms
would not be consistent within the perturbation theory employed. Such
higher derivative terms have been proved to be absent in spatially
flat isotropic models sourced by a free massless scalar
\cite{BouncePert}, but do arise in more general settings of physical
interest.

Secondly, one needs to ensure the robustness of results obtained by
employing models which take into account only a subset of such
corrections.  In particular, only results which can be shown to be
robust with respect to the addition of corrections which have been
ignored, as well as changes in the ambiguity parameters, can be
treated as realistic.  So far, on the other hand, robustness has most
often been studied within the more narrow settings of individual types
of corrections. An exception is the ongoing perturbative analysis of
\cite{BouncePot,QuantumBounce} as described in
Sec.~\ref{s:QuantState}.  This is particularly important in connection
with higher derivative corrections, whose dynamical consequences are
likely to be far more complicated than those corresponding to higher
order terms.  This is mainly due to the fact that such higher
derivative terms would change the character of the evolution equations
by changing their order. In fact ignoring such higher derivative terms
amounts to singular perturbations of the evolution equations, with
potentially important dynamical consequences.  Their physical
significance is given by capturing state properties. Only these
corrections provide a controlled generic implementation of quantum
aspects such as the spreading or deformations of states.

Thus the construction of realistic cosmological scenarios in loop
quantum cosmology will need to take into account all three types of
corrections. Particular attention needs to be paid to the question of
robustness of models which include these corrections only
partially. As an example we note that even though inverse volume and
higher order corrections have often been found to enhance one another,
there are specific settings where they compete against each other, such
as when a universe sourced by a free massless scalar enters the deep
quantum regime \cite{BouncePert}.

In the following section we shall summarise some of the
phenomenological scenarios constructed using various corrections.
Clearly, in view of the potential difficulties discussed above it is
important to bear in mind that details and even qualitative aspects of
such models may change once the complete set of quantum corrections
are taken into account.
\section{Non-singular and oscillating universes}
According to classical singularity theorems, gravitational collapse
generically results in a singularity
\cite{HawkingEllis,Senovilla}. These theorems are mainly statements in
differential geometry which use Einstein's equations only in one place,
in order to relate the Ricci curvature to the energy-momentum.  The
use of Einstein equations is nevertheless crucial since it allows the
geodesic behaviour, corresponding to trajectories of test masses, to
be related to the positive energy conditions. In particular for
ordinary matter and reasonable types of sources of energy encountered
classically, this would result in focusing of the geodesic families,
i.e.\ gravitational attraction in physical terms. It is this focusing
which is essential for the generic singularity theorems to hold. Thus
the singularity theorems rely both on Einstein's equations as well as
the energy conditions that are imposed.

\subsection{Bounces from inverse volume corrections}
\label{s:BounceInv}
The loop quantum corrections discussed above have implications for both
ingredients of the singularity theorems. 
Inverse volume corrections change the
matter Hamiltonian and thus can, effectively, break some of the
positive energy conditions in quantum gravity regimes. This easily
follows from the discussion in previous sections where we noted that 
the pressure corresponding to the
effective fluid becomes negative as we approach 
the singularity. (In some cases, the effective
fluid even behaves like a `phantom' fluid with an equation of state $w
< -1$.) This amounts to the violation of the null energy condition,
{\em independently of the form of the scalar potential $V(\phi)$},
thus removing an important obstacle to singularity avoidance -- but
without introducing non-physical concepts such as a negative kinetic
energy term which is otherwise used to model a `phantom fluid'
\cite{phantom}. Similarly, higher order and higher derivative terms
would change Einstein's equations themselves
even without considering the form of
matter. Although it is not as easy to see, these corrections can change the
relation between focusing and positive energy conditions
to the extent that even
classical matter satisfying all positive energy conditions can start
to de-focus geodesics once they enter a strong quantum regime. Thus,
just before a classical singularity could be reached, the dynamics of
space-time is changed in a way which allows it to evade the devastating 
consequences predicted by the classical singularity theorems.

Indeed, one can intuitively understand quantum effects as being
associated with the emergence of {\em repulsive contributions} to the
gravitational force on small scales, which would otherwise always be
attractive classically \cite{WS:MB}. This is again easier to see for
inverse volume corrections. We can view Eq.~(\ref{hamil}), taking for
simplicity the spatially flat case ($k=0$), as the energy equation of
a classical mechanical system with kinetic term $\dot{a}^2a$ and an
$a$-dependent potential
$W(a)=-\frac{1}{2}D(a)p_{\phi}^2/a^3-a^3V(\phi)$. For a complete
analysis we would have to know the dynamical behaviour of $\phi$.
This, however, could be avoided by assuming that the potential
$V(\phi)$ vanishes, which would imply that $p_{\phi}$ is constant.
Then, $W(a)\propto -D(a)/a^3$, which allows the gravitational force to
be readily determined as $F(a)=-W'(a)\propto -3D(a)/a^4+D'(a)/a^3$.
Classically $D(a)=1$ and the force is negative, pointing to smaller
volumes and thus implying attraction. As a result a classical
universe, once it starts collapsing, has no way of avoiding a
singularity since gravity only enhances the collapse. In a quantum
regime, however, $D(a)\not=1$ and in fact increases for small $a$. The
second term in the force is thus positive in these regimes and can
dominate the first term. Thus, the gravitational force becomes
repulsive and may prevent the total collapse (or drive an
inflationary, accelerated expansion phase). These effects thus allow
the possibility of a non-singular bounce for a collapsing universe
instead of a singular origin.

In fact the universe could be eternal.  To see this, consider a
positively curved isotropic universe satisfying Eq.~(\ref{hamil}) with
$k=1$. A bounce corresponds to a turning point (a minimum) of the
scale factor, treated as a function of time. For a bounce to occur,
there must thus exist a solution to Eq.~(\ref{hamil}) which at some
point satisfies $\dot{a}=0$ and $\ddot{a}>0$. This requires a balance
at this point between the right hand side of the equation and the
matter Hamiltonian on the left hand side.  Such a bounce is impossible
to occur in approaching a singularity within a classical framework
since in that case as $a\to 0$ the matter energy on the left hand side
diverges.  (Solutions to $\dot{a}=0$ are possible but have
$\ddot{a}<0$, thus corresponding to a recollapse.)  The effect of
including the correction term $D(q)$ on the left hand side of the
Eq.~(\ref{hamil}) is that on very small scales the energy density
decreases, even during the collapse phase, in such a way that it is
eventually balanced by the term involving $a$ on the right hand side.
Moreover, $\ddot{a}>0$ is then related to the presence of a negative
pressure.  (See \cite{BounceClosed,Oscill} for detailed
discussions. Similar bounces result even for spatially flat models
when the potential can become negative \cite{Cyclic}.)

Since positively curved universes can recollapse after an expanding
phase, the presence of the bounce allows oscillations, i.e.\ bounces
followed by recollapses, to occur. A transparent way of seeing this is
to employ a dynamical systems approach \cite{Oscill} (see also
\cite{BounceQualitative}). Assuming for simplicity a constant
potential, the effective evolution equations can in this case be
expressed as a two-dimensional autonomous system in the form
\cite{Oscill}
\begin{equation}
\dot{H}=\left( \frac{8 \pi G}{c^2} V -3H^2 \right)
\left( 1-\frac{A}{6} \right )
+ \frac{kc^2}{a^2} \left ( \frac{A}{2}-2\right )\quad,\quad
\dot{a}=Ha
\end{equation}
where $H =\dot a/a$ is the Hubble parameter, $A(a) \equiv {\rm d} \ln
D /{\rm d}\ln a$ and $V$ is the potential, assumed constant. The
equilibrium points of this system, obtained by setting the left hand
sides of both equations to zero, correspond to static solutions
defined by $\dot a = 0 = \ddot a$.  It can be shown that depending
upon the value of the potential $V$ there are up to two equilibrium
points for this system which correspond to saddle and centre
equilibrium points respectively \cite{EmergentLoop}.  It is
instructive to contrast this with the classical general relativistic
case where there is only one equilibrium point -- the \emph{Einstein
static universe} -- corresponding to a saddle point which is unstable.

\begin{figure}
\resizebox{0.7\textwidth}{!}{%
  \includegraphics{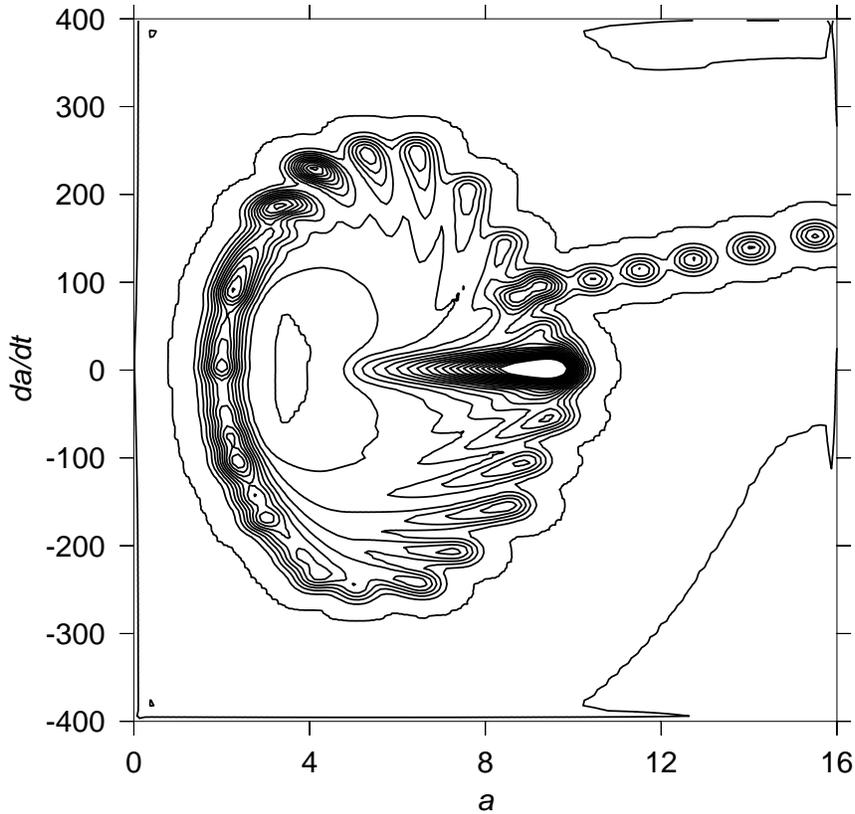}}
 \caption{\label{Phaseb} A stroboscopic phase diagram, showing the
density of discretized trajectories in the plane spanned by the scale
factor and its time derivative. The circular feature illustrates
oscillations around a stable equilibrium point near $a\sim 4$, ${\rm
d} a/{\rm d} t=0$ around which the universe performs a number
(possibly infinite) of small initial oscillations. At some point, it
comes close to a saddle point and escapes into an inflationary phase,
as indicated by the protruding linear feature.  }
\end{figure}

A further important finding in this connection is that the center
equilibrium point always occurs in the effective domain ($a<a_*$)
while in contrast the saddle equilibrium point always occurs in the
classical domain ($a>a_*$).  Thus the crucial consequence of the LQC
corrections is to induce an extra centre equilibrium point which is
capable of supporting oscillations in its neighbourhood and which
importantly always occurs in the effective domain.  An illustration of
the role of the center equilibrium point in determining the dynamics
is depicted in Fig. \ref{Phaseb}.  The circular orbits correspond to
oscillations around the centre equilibrium point, which eventually get
disrupted when the system enters an inflationary phase.

\subsection{Bounces from higher order corrections}
As was mentioned, if the matter source is a free scalar with a zero
potential and the universe is spatially flat ($k=0$), then there are
no higher derivative correction terms present and in that case the sum
of {\em all} higher order terms can be trusted to provide a consistent
effective equation \cite{BouncePert}. In that case, including all
higher order correction terms for the example (\ref{higher}), but
ignoring for now the inverse volume effects, the Friedmann equation
becomes
\begin{equation} \label{Fried1}
 \frac{8\pi G}{3c^4}\frac{p_{\phi}^2}{a^3}= \frac{a^3}{\ell_{\rm P}^2}
\left(1-\sqrt{1-4\frac{\ell_{\rm P}^2}{c^2}\frac{\dot{a}^2}{a^2}}\right)
\end{equation}
where $p_{\phi}$ is constant in the absence of a potential. Defining
the constant $C=4\pi G\ell_{\rm P}^2p_{\phi}^2/3c^4$ of the dimension
${\rm length}^6$, this equation can readily be solved to give
$\dot{a}=(c/\ell_{\rm P})a^{-5}C\sqrt{a^6/C-1}$.  In contrast to the
classical case, there exists a solution for $\dot{a}=0$ at a non-zero
value of the scale factor $a=C^{1/6}$. Moreover, we can take a further
time derivative of $\dot{a}$ and, after some simplifications, compute
$\ddot{a}= -5\dot{a}^2/a+3c^2C/a^5\ell_{\rm P}^2$.  This demonstrates
that at the point where $\dot{a}=0$ we have $\ddot{a}>0$, implying a
minimum -- or a bounce rather than a collapse. Thus a contracting
universe within this framework reaches a non-zero size before
re-expanding, never hitting a singularity.  This has been discussed in
\cite{EinsteinStaticHol} in an oscillatory context in parallel to what
is mentioned in Sec.~\ref{s:BounceInv}.

One can even compute precisely how a wave function of the universe
behaves, i.e.\ not only find solutions for expectation values such as
$a$ but also for fluctuations and other wave function parameters as
well \cite{BouncePert,BounceCohStates}. It turns out that they evolve
through the bounce unscathed, in such a way that the wave
function extends to a regime before the classical singularity. Detailed
pictures have been developed numerically for this model
\cite{QuantumBigBang,APS}, focusing on unsqueezed Gaussian states only
and large values of $p_{\phi}$. The results remain true if inverse
volume corrections are included, although details of the solutions
clearly change. This is an example of a model where different types
of quantum corrections have been included. In fact, the set of
corrections employed in this case is the complete set, since the
neglected higher derivative correction terms are exactly zero for this
model.

Despite this seeming robustness, however, small changes to the details
of the model, such as for example using an alternative choice of
factor ordering of the quantum operator determining the dynamics,
could render the higher derivative correction terms non-zero and hence
the model incomplete.  During the long evolution times for the
universe, such effects can easily add up and change the behaviour
considerably. Thus, the exact bounce solutions using only higher order
terms cannot easily be extended beyond the simplest model. For
example, the addition of a matter potential or the inclusion of
anisotropies and inhomogeneities, which constitute crucial ingredients
for a realistic cosmological model, could lead to strong quantum
back-reaction effects caused by the spread and deformation of the
quantum wave packet on its mean position. This highlights the need to
bear in mind the robustness of the models employed. The occurrence of
a bounce in this case could be prevented by quantum back-reaction
effects if they turned out to be substantial. While it still remains
to be seen whether or not this is the case, some qualitative results
are already known as described in the next subsection.  Prior to
knowing the outcome of such studies, therefore, scenarios of this type
while providing suggestive insights, cannot be guaranteed to be
robust.

\subsection{Role of the quantum gravity state}
\label{s:QuantState}

The crucial question to be understand before one can draw conclusions
about the robustness of scenarios in quantum cosmology is the quantum
back-reaction of an evolving state on its expectation values. Since
cosmology requires long time evolution, such quantum corrections which
are typically small but almost always present can add up to yield
significant contributions. In general, the presence of such
back-reaction effects implies that one is dealing with a highly
coupled system of equations which describes the time evolutions of the
infinitely many state parameters. This system of equations can be
solved exactly only in very rare cases, such as the free scalar system
available for cosmology \cite{BouncePert}, i.e.\ a free, massless
scalar in a spatially flat universe ($k=0$). As a free system, in
analogy with the harmonic oscillator in quantum mechanics or free
quantum field theories, this model can be used as the basis of a
perturbation expansion in interacting systems. In our context, this
can in particular be applied to scalar field models with non-vanishing
potentials, thus providing them with mass or self-interactions, or to
models with anisotropies and inhomogeneities. So far, the case of
a non-vanishing potential has been studied in some detail.

Thus, there are new correction terms which change the free solutions
and which can be derived and analyzed. It is particularly important to
reanalyze the bouncing scenarios in presence of these new correction
terms in order to make sure that earlier conclusions concerning the
existence of bounces are robust.  It is not just the presence of the
potential, which already changes the classical dynamics, but as seen
in the quantum mechanical example of Sec.~\ref{s:Higher} also
deviations from the free model which result in the spreading and other
features of a state to back-react on the effective evolution of
expectation values.  Thus in addition to inverse volume and higher
order corrections, corrections of the third type discussed above
become important.

Since the spreading itself is time dependent, there are new equations
in addition to (\ref{hamil}) for the dynamics of quantum variables
which do not have classical counterparts. Already in the free system,
this provides crucial information on the behavior of dynamical
coherent states showing how a state generically spreads as it
evolves. With such solutions, one can find relations between, e.g.,
fluctuations of a generic state before and after the bounce. The ratio
of these quantum variables turns out to be rather weakly restricted
\cite{BeforeBB,Harmonic}, such that even in the highly controlled free
system the pre-bounce state is not strongly determined by conditions
one may pose after the bounce. The degree of asymmetry of fluctuations
before and after the bounce is determined by a parameter $\eta$ which
defines the squeezing of the state of the universe.

For models with a self-interacting scalar field, the same spreading
effects happen but now they back-react on the behavior of expectation
values. Thus, Eq.~(\ref{hamil}) is replaced with an effective
Friedmann equation capturing this quantum back-reaction. This has the
form \cite{QuantumBounce,BounceSqueezed}
\begin{equation} \label{Fried2}
 \left(\frac{\dot{a}}{a}\right)^2 = \frac{8\pi G}{3c^2}\left(\rho
 \left(1-\frac{\rho_Q}{\rho_{\rm crit}}\right)
 \pm\frac{1}{2}\sqrt{1-\frac{\rho_Q}{\rho_{\rm crit}}} 
\eta V(\phi)+ \frac{a^6V(\phi)^2}{2p_{\phi}^2}\eta^2
\right)
\end{equation}
where $\rho_Q$ is an expression for energy density carrying quantum
corrections and $\rho_{\rm crit}=3c^4/8\pi G\mu^2$ is a critical
density defined in terms of a fundamental length parameter $\mu$ which
arises, as in the case of $a_*$, due to quantum gravity. (The previous
example (\ref{Fried1}) for a free scalar with $V(\phi)=0$, when solved
for $(\dot{a}/a)^2$, corresponds to $\mu=\ell_{\rm P}$. In general,
$\mu$ can differ from the Planck length by numerical factors but, more
importantly, can also depend on the scale factor $a$. The precise
behavior reflects the underlying class of quantum gravity states whose
evolution is described by these effective equations
\cite{InhomLattice,CosConst}.)

Moreover, $\eta$ in (\ref{Fried2}) is a parameter determining the
dynamical squeezing of the state.  Thus, the precise squeezing, or the
value of quantum correlations summarized by the parameter $\eta$, are
important for the fate of the bounce: when $\rho_Q\sim \rho_{\rm
crit}$, such that a free model would bounce at this point, with the
dominant term on the right hand side depends on $\eta$. Now since this
is not a constant but a dynamical variable which changes in time; it
is generically non-zero at the would-be bounce. Consequently, the
condition $\rho_Q\sim \rho_{\rm crit}$ would not result in a bounce
because in this case $\dot{a}\not=0$. Whether or not there will always
exist a time time at which $\dot{a}$ vanishes is yet to be determined from
an analysis of the coupled equations of motion which is still in
progress. Only if this is ensured can one conclude the robustness of
the bounce.

\section{Complete universe scenarios: Combining conceptual and
observational issues}
As was mentioned above, despite its many recent successes, one of the
major shortcomings of general relativistic standard cosmology is the
presence of an initial singularity at which the classical laws of
physics break down and with it the classical notion of space-time,
causality, and in effect scientific predictability.  Thus the question
of the origin of the universe cannot be addressed within the classical
framework.  In the above sections we have briefly discussed the
general outlines of how quantum gravity effects may change the
classical picture and resolve this dilemma.

In addition to the question of origin, the other central question in
standard cosmology is how to situate the early inflationary phase of
the universe within a fundamental theory of quantum gravity.
According to current cosmological understanding such a phase is
crucial for structure formation.  Thus clearly a viable cosmological
scenario needs to deal with both questions: providing mechanisms to
remove classical singularities as well as initiating a successful
phase of inflation (or an alternative to provide structure formation).

In the above discussions we have already alluded to this possibility,
particularly since the presence of an inflationary phase is likely to
be closely related to the presence of bounces, given that both require
the condition $\ddot{a}>0$ for their occurrence.  A model successfully
answering both these questions may provide the fascinating prospect of
indirectly testing the bounce phase through precise observations of
background radiations.

Clearly much work remains to be done, both theoretically and
observationally, in order to be in
a position to test such a possibility.
Nevertheless, the recent developments make it
at least possible to study whether the existence
of cosmologically viable scenarios of this type is plausible 
within the LQC framework.
\subsection{An emergent non-singular universe setting the initial
conditions for inflation}
Attempts to construct oscillating cosmological models go back a number
of decades \cite{Tolman}. These clearly predate the recent attempts to
employ specific quantum gravity effects in order to avoid a singular
origin for the universe.  An interesting example of such an attempt is
the so called emergent scenario according to which the universe is
past eternal by being asymptotic to an Einstein static universe
\cite{Emergent}. This provides an example
of a non-singular cosmological model within the general relativistic
framework.  Despite its novelty, the main shortcoming of this scenario
is that, as we noted above, within the general relativistic framework,
the classical Einstein static universe corresponds to a saddle
equilibrium point which is unstable to all homogeneous-isotropic
perturbations, including quantum fluctuations. (Interestingly though
this solution has been shown to be stable with respect to
inhomogeneous perturbations \cite{StableEinsteinStatic}.) This
therefore prevents such a model from being past eternal, unless the
universe is initially at the equilibrium point, which amounts to
extreme fine tuning.

As we saw above the situation is radically different 
when LQC inverse volume corrections 
are included. In the semi-classical domain
the appropriate equilibrium point turns out to be a 
centre. An important consequence of this change in the nature
of the equilibrium point is that small perturbations, which
are unavoidably present, do not lead to a radical
departure from the equilibrium state, but result in
oscillations about it. This allows the construction of a plausible
emergent scenario with a non-singular origin for the universe
\cite{EmergentLoop,EmergentNat}.

\begin{figure}
\resizebox{0.65\textwidth}{!}{%
  \includegraphics{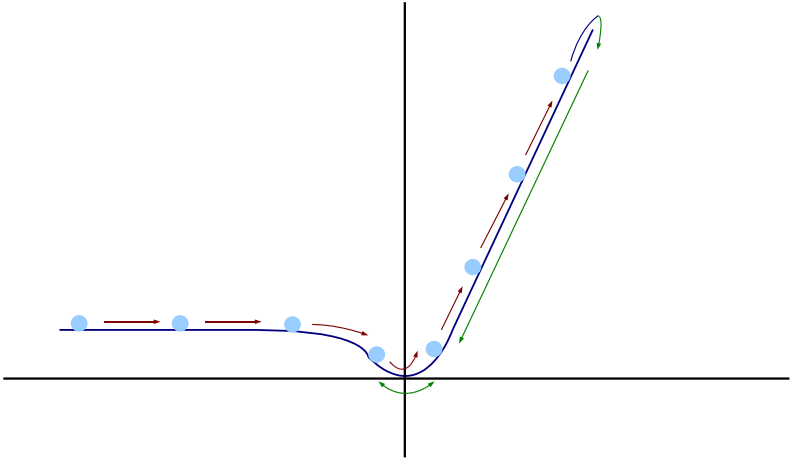}}
\caption{\label{Sketch} {A sketch illustrating how the oscillations of
the universe push the field value (horizontal) to the right and
eventually up the potential (vertical), setting the required initial
conditions for the onset of slow-roll inflation.  During inflation the
field rolls back down the potential and eventually oscillates about
the minimum and decays.  }}
\end{figure}

As an example of such a scenario, consider a positively curved FLRW
universe sourced by a scalar field with a potential which
qualitatively has the form given in Fig.~\ref{Sketch}. Such potentials
have been shown to arise in a number of settings of interest,
including in attempts at renormalization of theories of quantum
gravity \cite{HigherDerivUnit}, as well as in low-energy limits of
superstring theories \cite{VacuumSuperStrings}.  Now let the field be
initially placed far away to the left along the $\phi$-axis
(i.e. $\phi \to - \infty$) with $\dot{\phi}_{\rm in} >0$ 
and let the universe be close to the centre equilibrium point and
hence oscillating about it. A crucial consequence of the oscillations
of the universe is to push the field uni-directionally to the right
along the $\phi$-axis and eventually up the potential to the point
where $\phi$ is such that $V(\phi) \approx \dot{\phi}_{\rm in}^2/2$
at which point the oscillations stop (see \cite{Oscill,EmergentLoop}
for the details). The field then turns round and the slow-roll
inflation begins, finally leading to oscillations about a minimum of
the potential and subsequent reheating at the start of the standard
hot big bang phase. Importantly, it has been shown that such
oscillations can push the field high enough up the potential to
successfully set the initial conditions for the onset of inflation
\cite{EmergentLoop} (see also \cite{LoopFluid} for an analysis in the
presence of other matter sources). This allows the universe to
subsequently undergo a sufficiently large number of $e$-foldings
required by observations.
\subsection{Higher order corrections and oscillations}
The above oscillatory scenario employs inverse volume
corrections. Such models have also been discussed for higher order
corrections to construct oscillating universe models
\cite{svv,EinsteinStaticHol}. These models are based on subseries of
higher order terms in $\dot a$ with arbitrarily high powers. Still,
these are not the complete equations which would be more difficult to
obtain. This is because such oscillating models necessarily deviate
from the exactly solvable one of a free scalar in a spatially flat
universe, requiring non-zero potentials. As a result the inclusion of
such an entire subseries of higher order terms is not consistent with
ignoring the higher derivative correction terms.  For example, in
presence of a scalar potential, as in the case of the model considered
in \cite{svv}, a back-reaction occurs between the fluctuations of a
quantum state and its expectation value, implying that for the model
to be consistent further corrections need to be taken into account as
in Sec.~\ref{s:QuantState}.  Almost all the higher order terms are
exceedingly small compared to other corrections, and yet they are all
added up.  If one were to treat higher order terms as an approximation
to the full equations in the presence of quantum back-reaction,
self-consistency of the perturbative analysis would require the
inclusion of only a finite number of higher order correction terms,
rather than the full set. (In the variables introduced in
\cite{BouncePert}, on the other hand, one can essentially perform a
resummation to ensure this consistency.)  This gives rise to equations
which are noticeably different from those that have so far been used
in such models.

Oscillations also arise if one keeps the scalar field massless and
non-interacting but allows positive spatial curvature, as was
numerically studied in \cite{APSCurved} for initial states which are
unsqueezed. Although back-reaction still occurs, it is only active
briefly at each recollapse since the bounce phases are still well
described by the solvable model. There is thus not much spread of the
quantum state unless one considers many cycles of bounces and
recollapses. In the presence of a scalar potential, however,
back-reaction especially of correlations can be so strong that it
might even prevent a single bounce from occurring. This example
demonstrates that at present also scenarios based on higher order need
to be treated with great care. 
Methods for a complete treatment including all quantum corrections in
isotropic models are now available and being applied to oscillating
scenarios \cite{BouncePot,QuantumBounce}.

\section{Conclusions}
We have seen that loop quantum gravity gives rise to three different
types of corrections in the effective equations: inverse volume
corrections, higher order terms in $\dot a$ and quantum back-reaction,
related to higher time derivative terms.  The first two are typical
for quantum geometry effects as they arise from loop quantum gravity,
while the latter are of genuine quantum origin.  At present the
complete set of corrections is known precisely only for the case of a
free massless scalar in a flat isotropic geometry
\cite{BouncePert,BounceCohStates}.  More general settings are then
treated by perturbations around the solvable model, much like
interacting quantum field theories are dealt with by perturbations
around a free theory.  in more general settings it still remains to be
determined. However, as these developments are quite recent, and still
ongoing, most models considered so far include only a subset of
corrections that are expected to be present. The overlooked
corrections, however, specially the higher derivative corrections, can
potentially change the behaviour of the models dramatically.  Although
each correction is typically small, long evolution times are by
definition involved in cosmological scenarios, where even small terms
can have lasting effects.  In view of these limitations, it is
therefore important to pay particular attention to the questions of
consistency and robustness of models employed, if their predictions
are to be treated as wholly reliable.

Despite these shortcomings, however, recent developments in LQC have
provided a framework which in principle allows the systematic
derivation, presently in progress, and testing of all relevant quantum
corrections. In yet more complicated models quantum dynamics has been
analyzed in regimes where effective equations break down and are
replaced by difference equations for the evolution of the quantum
states which again have been shown to remain non-singular
\cite{Sing,HomCosmo,Spin,SphSymmSing}. These difference equations are
capable of allowing the evolution of a state through strong quantum
regimes where no effective description is valid.  In fact, for general
solutions to a difference equation it is easily possible for the
quantum fluctuations to remain strong even after evolving through a
classical singularity. Thus, although the quantum state would stay
well-defined, it may not easily allow a semiclassical geometry after a
``bounce.''  Even recent studies of the effective equations indicate
that fluctuations generically differ on both sides of the bounce
\cite{BeforeBB,Harmonic}.  An emergence of a classical universe after
a bounce could then only be understood in more complicated terms,
based e.g.\ on decoherence \cite{Decoherence}.  The difficulty with
employing difference equations for a quantum state rather than an
effective geometry, however, is that they are less intuitive and not
easily conducive to the development of cosmological scenarios.

Despite the shortcomings mentioned above, the development of
preliminary scenarios, using partial inclusion of corrections, have
nevertheless been valuable.  In particular, the emergent universe
scenario sketched here, based on effective equations, provides an
example of how LQC effects can potentially give rise to physically
reasonable non-singular behaviour.  The important features of this
scenario are that it is non-singular, past-eternal, and
oscillating. Importantly, the oscillations have a crucial function,
providing a mechanism for setting the initial conditions for
successful inflation. Furthermore, this scenario is in principle
observationally testable.  It sets the initial conditions for the
standard hot big bang expansion which is broadly supported by current
astronomical observations, but requires a positively curved
universe. Thus although this scenario is at present compatible with
observations, it can in principle be ruled out, especially given the
planned future observations with far greater resolutions. Also, the
turn over of the scalar field can potentially leave signatures in the
CMB which may be observable \cite{InflationWMAP}. Finally,
inhomogeneous perturbations in effective equations
\cite{HamPerturb,QuantCorrPert,vector,tensor} may leave additional
observable imprints
\cite{ScaleInvLQC,InhomEvolve,CCPowerLoop,SuperInflLQC}.  It should
also be mentioned here that a number of other recent early universe
models postulate a bounce (e.g.\ \cite{CyclicEkpy}), but
none in a fully satisfactory way within the context of a theory of
quantum gravity and without introducing other singularities.

The above discussion has demonstrated the potential
of LQC to provide a unified framework to deal with the questions of origin
and the early inflationary phase of the universe.
Despite these preliminary successes, however,
a great deal of work remains to be done.
In particular, the construction of consistent and robust realistic scenarios
needs to await the evaluation of the complete set
of corrections which are expected to be present.


\bigskip

\noindent {\bf Acknowledgements:} 
This work was supported in part by NSF grants PHY0554771 and PHY0653127.


\end{document}